\documentclass[epj-spec]{svjour}
\usepackage[dvips]{graphicx}
\usepackage{graphics}
\usepackage{amsmath}
\usepackage{amsfonts}
\usepackage{amssymb}
\usepackage{cite}

\newcommand{\goesto}{\rightarrow}
\newcommand{\bra}[1]{{\langle #1 |}}
\newcommand{\ket}[1]{{| #1 \rangle}}

\newcommand{\ave}[1]{{\langle #1\rangle}}
\newcommand{\expec}[1]{{\mathbb E}(#1)}
\providecommand*{\dd}{\mathrm{d}}
\providecommand*{\dod}[2]{\frac{\dd #1}{\dd #2}}

\providecommand*{\Tr}{\operatorname{Tr}}
\providecommand*{\Kom}[2]{[#1,#2]}
\providecommand*{\Ket}[1]{\left|#1\right>}
\providecommand*{\Bra}[1]{\left<#1\right|}

\begin{document}

\title{Heat Transport in Quantum Spin Chains}
\subtitle{Stochastic Baths vs Quantum Trajectories}
\author{Carlos Mejia-Monasterio\inst{1}\fnmsep\thanks{\email{mejia@calvino.polito.it}} \and Hannu Wichterich\inst{2}\fnmsep\thanks{\email{hwichter@uos.de}}}
\institute{Dipartimento di Matematica, Politecnico di Torino, Corso Duca degli
  Abruzzi  24 I-10129  Torino,  Italy. \and  Fachbereich Physik,  Universit\"at
  Osnabr\"uck, Barbarastrasse 7, D-49069 Osnabr\"uck, Germany.}

\abstract{We  discuss the  problem of  heat conduction  in quantum  spin chain
  models.  To investigate this problem  it is necessary to consider the finite
  open system connected to heat baths. We describe two different procedures to
  couple the system with the reservoirs:  a model of stochastic heat baths and
  the  quantum trajectories  solution  of the  quantum  master equation.   The
  stochastic heat  bath procedure  operates on the  pure wave function  of the
  isolated  system, so  that it  is locally  and periodically  collapsed  to a
  quantum state consistent with  a boundary nonequilibrium state. In contrast,
  the quantum  trajectories procedure evaluates ensemble averages  in terms of
  the reduced density matrix operator of the system. We apply these procedures
  to  different models  of  quantum  spin chains  and  numerically show  their
  applicability to study the heat flow.}

\maketitle

\section{Introduction}
\label{intro}

The  derivation of  Fourier's  law  of heat  conduction  from the  microscopic
dynamics,  without any  ad hoc  statistical assumption,  is one  of  the great
challenges  of  nonequilibrium  statistical  mechanics  \cite{bonetto}.   This
problem, in spite of having a long history, is not completely settled: Given a
particular classical,  many-body Hamiltonian system,  neither phenomenological
nor  fundamental transport  theory can  predict whether  or not  this specific
Hamiltonian  system yields  an energy  transport governed  by the  Fourier law
$J=-\kappa\nabla  T$, relating the  macroscopic heat  flux to  the temperature
gradient $\nabla T$ \cite{peierls}.

%Heat flow is  universally presumed to obey a  simple diffusion equation, which
%can  be regarded  as the  continuum limit  of a  discrete random  walk.   As a
%consequence,  transport  theory  requires  that the  underlying  deterministic
%dynamics  yield  a truly  random  process.  Therefore,  it  is  not mere  idle
%curiosity  to wonder  what class,  if any,  of many-body  systems  satisfy the
%necessary   stringent   requirements.

At the  classical level  a great amount  of work  has focused on  the relation
between  the chaotic  character of  the  microscopic dynamics  and the  normal
macroscopic                                                           transport
\cite{lebowitz,ford,casati85,prosen,leprifpu,LLP,alonso,tria,dhar01,aoki01,grassberger,mejia-1,mejia-2,li,EY,EMMZ},
(see also Ref.~\cite{LLP-review}  for a recent review ).   The general picture
that emerges from these investigations  is that deterministic chaos appears to
be an  essential ingredient required  by transport theory.   However, strictly
speaking, the exponential instability  that characterizes the chaotic dynamics
is  neither sufficient  \cite{leprifpu} nor  necessary \cite{tria,li}  for the
validity of  Fourier law. What  has been shown  in Ref.~\cite{li} is  that the
diffusive behavior,  which is  at the  root of normal  heat transport,  can be
obtained even  if the Lyapunov exponents  are zero. This  constitutes a strong
suggestion that normal heat conduction  can take place even without the strong
requirement of exponential instability.

At the quantum level, the question whether normal transport may arise from the
underlying     quantum      dynamics     is     a      controversial     issue
\cite{zotos,saito00,MHGM,saito03,MMG,MMPC,jung}. This is  mostly because it is
not clear how  to describe the transport of energy or  heat from a microscopic
point of view.   In analogy to classical systems, a  quantum derivation of the
Fourier's law calls directly in question the issue of quantum chaos \cite{qc}.
However, a main feature of quantum motion is the lack of exponential dynamical
instability  \cite{casati}.  This  fact   may  render  very  questionable  the
possibility to derive the Fourier law of heat conduction in quantum mechanics.
Thus it is  interesting to inquire if, and under  what conditions, Fourier law
emerges  from the  laws  of quantum  mechanics  (for a  recent  review of  the
microscopic foundations of the quantum Fourier's law see \cite{qfl-rev}).

\begin{figure}[!t]
\begin{center}
\includegraphics[scale=0.6]{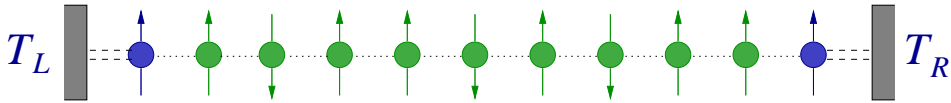}
\caption{Schematic representation of a finite one-dimensional quantum spin
  chain, coupled to external heat reservoirs at different temperatures.
\label{fig:chain}}
\end{center}
\end{figure}

To investigate  the problem of quantum heat  transport one has to  deal with a
finite open system  connected to heat baths. This fact  renders the problem to
obtain a derivation of the  quantum Fourier's law extraordinary difficult.  In
this paper we  present two complementary methods to study  the problem of heat
conduction in  quantum spin chain models. We  consider one-dimensional quantum
spin chain  models like the  one schematically shown in  Fig.~\ref{fig:chain}. 
Each spin  in the  chain interacts  with its neighbours  and possibly  with an
external  field.  The  range  and  type  of interaction  does  not  limit  the
applicability of our methods.  Furthermore,  the spin chain is in contact with
two heat reservoirs.  The Hamiltonian of the system can be written as
\begin{equation} \label{eq:Hgen}
{\mathcal H} = H_{system} + H_{reservoirs}(T_L,T_R) + H_{int} \ ,
\end{equation}
where  $H_{int}$ represents  the  coupling  between the  system  and the  heat
reservoirs.

In recent  years the  transport of heat  in quantum  spin chains has  been the
subject of intense experimental \cite{spin-exp} and theoretical investigations
\cite{zotos,castella,shimsoni,cabra,zotos04,saito00,MHGM,saito03,MMG,MMPC,jung,Wichterich2007}. 
In analogy to classical systems it  has been found that non integrable quantum
spin  chains show  normal  heat  transport, while  integrable  chains lead  to
ballistic transport and thus,  to divergent heat conductivity.  In particular,
in Ref.~\cite{MMPC} the validity of  the quantum Fourier's law has been linked
to  the onset  of quantum  chaos, in  the statistical  sense of  Random Matrix
Theory.  Moreover, in Ref.~\cite{MMPC} an intermediate case for which the spin
chain  is neither  integrable nor  chaotic was  shown to  have  divergent heat
conductivity.   Up to  our knowledge,  this is  the only  example for  which a
quantum  spin chain  with intermediate  statistics leads  to an  abnormal heat
transport  (see also  \cite{jung}  where almost  integrable  models have  been
studied).  In spite of the general evidence, in Ref.~\cite{MHGM} the Fourier's
law was numerically confirmed for a class of integrable, albeit small, quantum
spin chains.

In  these   investigations,  particularly   those  that  are   numerical,  the
thermodynamical limit  at which the  heat conductivity is formally  defined is
difficult, if  not impossible to  address. This limitation is  emphasized when
one deals  with non  integrable systems for  which a  theoretical perturbative
analysis is  not possible.  So far  the most popular  theoretical framework to
study heat transport is the Green-Kubo formula \cite{GK}, derived on the basis
of linear response theory.  However, the  use of GK formula for heat transport
requires  {\it ad-hoc}  statistical  assumptions that  lack  of a  microscopic
justification of  its applicability (see  {\it e.g.}, \cite{qfl-rev}  and also
\cite{GSM}  for  a plausible  derivation  of  a  Green-Kubo formula  for  heat
transport).   A second  standard  treatment  is based  on  the Quantum  Master
Equation (QME). However,  this method involves the calculation  of the reduced
density  matrix  of the  system,  thus  limiting  numerical investigations  to
relatively small system sizes.

Given this  state of  affairs, the development  of numerical methods  that can
deal with both integrable and non integrable spin models and that are amenable
to study large system sizes is very desirable. In the present paper we present
two  methods recently introduced  to study  the transport  of heat  in quantum
systems.  In section  \ref{sec:sto}, we discuss a model  of stochastic quantum
heat  baths  where $H_{int}$  consists  of  a  stochastic time-periodic  local
perturbation \cite{MMPC}.  This perturbative term acts on the, otherwise, pure
state of  the system.   We show that  this procedure  leads to a  well defined
local  temperature and  discuss the  validity of  Fourier's law  in  a Quantum
spin-$1/2$  Ising chain  in a  tilted  external magnetic  field.  This  method
allows  to study  finite  but large  spin  chains as  its implementation  only
requires  the  knowledge  of  the   pure  vector  state  of  the  system.   In
Section~\ref{sec:qme} we present a QME in Lindblad form that can appropriately
describe nonequilibrium steady  states \cite{Wichterich2007}.  Furthermore, we
use the Monte Carlo wave function formalism to study the heat transport in the
spin-$1/2$ Heisenberg  chain.  In  Section~\ref{sec:disc} a discussion  of the
results and applicability of our integration methods is presented, followed in
Section~\ref{sec:final} by our final remarks.

\section{Stochastic Quantum Heat Baths}
\label{sec:sto}

In  this Section  we  describe the  implementation  of a  stochastic model  of
quantum heat baths.

We  consider a  one-dimensional quantum  spin-$1/2$ chain  of length  $N$. The
Hamiltonian of the  system can be written in general  as in Eq.~\ref{eq:Hgen}. 
Each  spin in  the  chain is  coupled to  its  neighbours and  possibly to  an
external field. Furthermore,  let us consider that the  leftmost ($s_{0}$) and
rightmost ($s_{N-1}$) spins are coupled  via the $H_{int}$ term, with external
ideal heat  baths at temperatures  $T_L$ and $T_R$ respectively.   A schematic
representation of this general model is shown in Fig.~\ref{fig:chain}.

The aim  of this procedure is  to focus on the  evolution of the  state of the
spin  chain,  avoiding  to  introduce   any  particular  model  for  the  heat
reservoirs. Therefore, we  only assume that the heat baths  act locally on the
state of the respective spin so that the state of $s_{0}$ and of $s_{N-1}$ are
thermal states at the respective  temperatures.  This model for the reservoirs
is  analogous   to  the  stochastic  thermal  reservoirs   used  in  classical
simulations\cite{mejia-2} and we thus  call it a \emph{stochastic quantum heat
  baths}.

To be precise, we work in the representation basis of $\sigma^z$. Furthermore,
we  use  units in  which  Planck  and Boltzmann  constants  are  set to  unity
$\hbar=k_{\rm B}=1$ In this the wave function at time $t$ can be written as
\begin{equation}
\ket{\psi(t)}=\!\!\!\!\sum_{s_0,s_1,\ldots,s_{N-1}}\!\!\!\!\!\! C_{s_0,s_1,\ldots,s_{N-1}}(t)
\ket{s_0,s_1\ldots s_{N-1}}\ ,
\end{equation}
where $s_n  =0,1$ represents  the \emph{up}, \emph{down}  state of  the $n$-th
spin,  respectively.  The  wave  function at  time  $t$ is  obtained from  the
unitary  evolution  operator   $\mathrm{U}(t)  =  \exp(-i\mathcal{H}t)$.   The
interaction  with the  reservoir is  not included  in the  unitary  evolution. 
Instead, we  assume that  the spin  chain and the  reservoir interact  only at
discrete times with  period $\tau$ at which the states of  the spins $s_0$ and
$s_{N-1}$ are  stochastically reset. The  evolution of the wave  function from
time $t$ to time $t+\tau$ can be represented as
\begin{equation} \label{eq:evol}
\ket{\psi(t+\tau)} = \Xi(T_L,T_R)\mathrm{U}(\tau)\ket{\psi(t)} \ ,
\end{equation}
where  $\Xi(T_L,T_R)$   represents  the  unitary  stochastic   action  of  the
interaction  with the  left and  right  reservoirs at  temperatures $T_L$  and
$T_R$ respectively.

The action of $\Xi(T_L,T_R)$ takes place in two steps:

\noindent   ({\it    i})
A local  measurement of the state  of the spins  coupled to the heat  baths is
performed.  Then  their state collapses to a  state ($s_0^*$,$s_{L-1}^*$) with
probability
\begin{equation}
p(s_0^*,s_{N-1}^*)=\sum_{s_1,\ldots,s_{N-2}}
|C_{s_0^*,s_1,\ldots,s_{N-2},s_{N-1}^*}|^2 \ .
\end{equation}
Numerically     this     means     that     we    put     all     coefficients
$C_{s_0,s_1,\ldots,s_{N-1}}$  with $(s_0,s_{N-1})  \neq  (s_0^*,s_{N-1}^*)$ to
zero. Afterwards, the wave function is renormalized.

\noindent  ({\it ii})
The new state of the edge  spins is stochastically chosen: $s_0$ and $s_{N-1}$
are set to {\em down},  ({\em up}) state with probability $\mu$,($1-\mu$). The
probability $\mu(\beta)$ depends  on the canonical temperature of  each of the
thermal reservoirs:
\begin{equation} \label{eq:canonical}
\mu(\beta)  =   \frac{e^{-\beta E_{down}}}{e^{-\beta E_{down}} + e^{-\beta E_{up}}} \ ,
\end{equation}
where $\beta  = 1/T$  is the inverse  temperature. This simulates  the thermal
interaction with the reservoirs.

This interaction thus  (periodically) resets the value of  the local energy of
the  spins  in  contact  with   the  reservoirs.   This  information  is  then
transmitted  to the  rest of  the system  during its  dynamical  evolution and
relaxation towards  equilibrium. Therefore, the  value of $\tau$  controls the
strength  of the coupling  to the  bath.  We  have found  that, in  our units,
$\tau=1$  provides an optimal  choice.  On  one hand,  large values  of $\tau$
correspond to  a ``microcanonical'' situation:  if $\tau$ is much  larger than
the relaxation  time of the system then  the spin chain behave  as an isolated
system, leading to a equidistributed mean  energy. On the other hand, a strong
coupling with the reservoirs, {\it i.e.}, $\tau\ll 1$, the state of the system
freezes due  to the  quasi-continuous monitoring, a  situation similar  to the
Zeno  effect.  We  have checked  that intermediate  values of  $\tau$  lead to
qualitatively similar results.

Also, note that our method does not correspond to a stochastic unraveling of a
QME.  Here  the evolution of  the system between two  consecutive interactions
with  the baths  is  purely unitary  and not  dissipative  as for  the QME  in
Section~\ref{sec:qme}.

The  use of  stochastic quantum  heat  baths has  the advantage  that it  only
requires the  calculation of the vector  state of the  system (\ref{eq:evol}). 
This allows  to computed time averages  for spin chains longer  than the sizes
that can be studied with other methods. 

Finally,  we remark that  the stochastic  quantum heat  baths method  does not
depend on the range and type of the interaction.

\subsection{Fourier's law in the Ising chain in a tilted magnetic field}

In this Section  we discuss the heat  transport in a Ising chain  of $N$ spins
$1/2$ with coupling constant $Q$  subject to a uniform magnetic field $\vec{h}
= (h_x,0,h_z)$, with open boundaries. The Hamiltonian reads
\begin{equation} \label{eq:H}
\mathcal{H} = \sum_{n=0}^{L-2}H_n +
\frac{h}{2}(\sigma_\mathrm{l} + \sigma_\mathrm{r}) \ .
\end{equation}
where $H_n$ are local energy density
operators
\begin{equation} \label{eq:H_local}
H_{n}     =   -Q\sigma^z_n\sigma^z_{n+1}    +
\frac{\vec{h}}{2} \cdot \left(\vec{\sigma}_n  + \vec{\sigma}_{n+1}\right) \ ,
\end{equation}
and  $\sigma_\mathrm{l} = \vec{h}\cdot\vec{\sigma}_0/h$,  $\sigma_\mathrm{r} =
\vec{h}\cdot\vec{\sigma}_{N-1}/h$ are  the spin operators  along the direction
of  the magnetic  field of  $s_0$ and  $s_{N-1}$ respectively.   The operators
$\vec{\sigma}_n =  (\sigma^x_n,\sigma^y_n,\sigma^z_n)$ are the  Pauli matrices
for the $n$-th  spin, $n=0,1,\ldots N-1$. The direction  of the magnetic field
affects the qualitative  behavior of the system: it  is integrable for $h_z=0$
and not integrable otherwise. When $h_z$ is of the same order of $h_x$ quantum
chaos sets in \cite{MMPC}.  Using  the stochastic quantum heat bath formalism,
in Ref.~\cite{MMPC}  it was shown  that the validity  of the Fourier's  law is
related to the onset of quantum chaos.  In what follows, we consider a chaotic
chain (with  parameters $Q=2$, $h_x=3.375$,  $h_y=0$ and $h_z=2$)  and discuss
the establishment of local thermal  equilibrium in terms of the density matrix
operator.  Furthermore,  we also  recall  some  of  the results  presented  in
Ref.~\cite{MMPC} concerning the validity of the Fourier's law.

In order  to apply the stochastic formalism  to this model one  needs first to
rotate the  state of  the edge spins  to the  direction of the  external field
$\vec{h}$,  so  that  the  local  measurement described  above  ({\it  i})  is
meaningful.   This  is  done  by  rotating  the wave  function  by  the  angle
\mbox{$\alpha  =  \tan^{-1}(h_x/h_z)$} to  the  eigenbasis  of the  components
$\sigma_\mathrm{l}$ and  $\sigma_\mathrm{r}$, that is  $\ket{\psi} \rightarrow
e^{-i\alpha(\sigma_0^y+\sigma_{L-1}^y)/2}\ket{\psi}$.   After  the  stochastic
reset of  the edge spins the wave  function is rotated back  to the $\sigma^z$
basis,                         $\ket{\psi}                         \rightarrow
e^{i\alpha(\sigma_0^y+\sigma_{L-1}^y)/2}\ket{\psi}$.

To integrate the unitary evolution $U(t)\ket{\psi}$ of the system we have used
an  accurate high  order  split-step factorization  of  the unitary  evolution
operator  described  in Ref.~\cite{split-step}.   We  then consider  different
random  quantum trajectories  of  the randomly  chosen  initial wave  function
$\ket{\psi(0)}$ of the system.  The  state is then evolved for some relaxation
time $\tau_{\rm rel}$  after which it is assumed to  fluctuate around a unique
steady  state.   Measurements are  then  performed  as  time averages  of  the
expectation  value   of  suitable  observables.   We   further  average  these
quantities over the ensemble of ``quantum trajectories''.

\begin{figure}[!t]
\begin{center}
\includegraphics[scale=0.85]{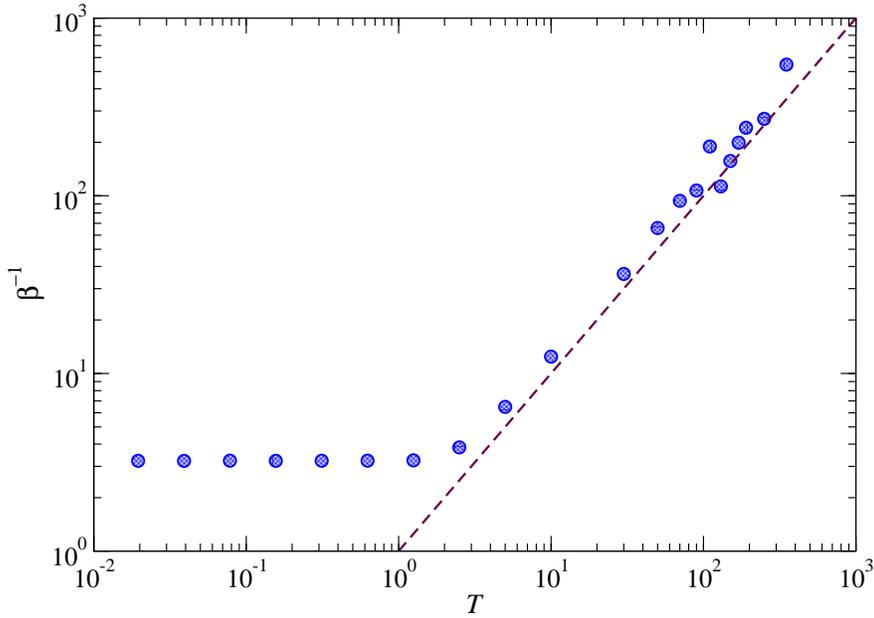}
\caption{
  Effective temperature at the bulk  $\beta^{-1}$ as a function of the nominal
  value of the temperature of the  baths $T$. $\beta$ was obtained from a best
  fit to a  exponential of $\rho_n(E_n)$ in the central  symmetry band for the
  chaotic chain of length $N=7$. The dashed line corresponds to the identity.
\label{fig:beta}}
\end{center}
\end{figure}

In  order to test  the effectiveness  of the  coupling between  the stochastic
quantum  baths and  the system,  we have  computed the  time  averaged density
matrix of the system
\begin{equation} \label{eq:rho}
\overline{\rho} = \lim_{t\goesto\infty} \int_0^t 
\ket{\psi(s)}\bra{\psi(s)}{\mathrm d}s \ ,
\end{equation}
where $\psi(s)$  is the state of  the system at time  $s$. Quantum statistical
mechanics postulates that if a system, described by a Hamiltonian $H$, is in a
thermal equilibrium state at temperature $\beta^{-1}$ then in the energy basis
($H\ket{\phi_n}=E_n\ket{\phi_n}$), the density matrix operator is
\begin{equation} \label{eq:rhoeq}
\bra{\phi_n}\rho\ket{\phi_m} = \frac{e^{-\beta E_n}}{\mathcal{Z}}
\delta_{m,n} \ ,
\end{equation}
where  $\mathcal{Z}  =  \sum_n  e^{-\beta  E_n}$ is  the  canonical  partition
function.

We  have  performed  equilibrium  simulations,  {\it  i.e.},  $T_L=T_R=T$  and
computed   the   time  averaged   density   matrix.    We   have  found   that
$\overline{\rho}$ is diagonal in within numerical accuracy.  Moreover, $\rho_n
\equiv \overline{\bra{\phi_n}\rho\ket{\phi_n}}$ is  an exponential function of
$E_n$ inside  each symmetry band.  This verifies  Eq.~\ref{eq:rhoeq} and thus,
that the system  reach a canonical equilibrium.  Furthermore,  from a best fit
to a  exponential of $\rho_n(E_n)$ one  can extract a value  for the effective
temperature at  the bulk  of the system.   In Fig.~\ref{fig:beta} we  show the
obtained effective bulk temperature $\beta^{-1}$  as a function of the nominal
value  of the  bath's  temperature  $T$.  At  low  temperatures the  effective
temperature  at the  bulk saturates  to a  constant which,  together  with the
energy profile $E_n$, is in principle determined by the ground state. However,
we have  found that the  saturation value does  not correspond to  neither the
ground state of  the two-body energy density operator $H_n$  nor to the ground
state of the many-body  operator $\mathcal{H}$.  At high temperatures ($T>5$),
the spin chain thermalizes to exactly  the same temperature as that set by the
stochastic heat  baths. This strongly  supports the effectiveness of  our bath
model.  Moreover,  this also  suggests the validity  of the  ergodic property,
namely  the  averages over  the  canonical  ensemble  are equivalent  to  time
averages.

\begin{figure}[!t]
\begin{center}
\includegraphics[scale=0.85]{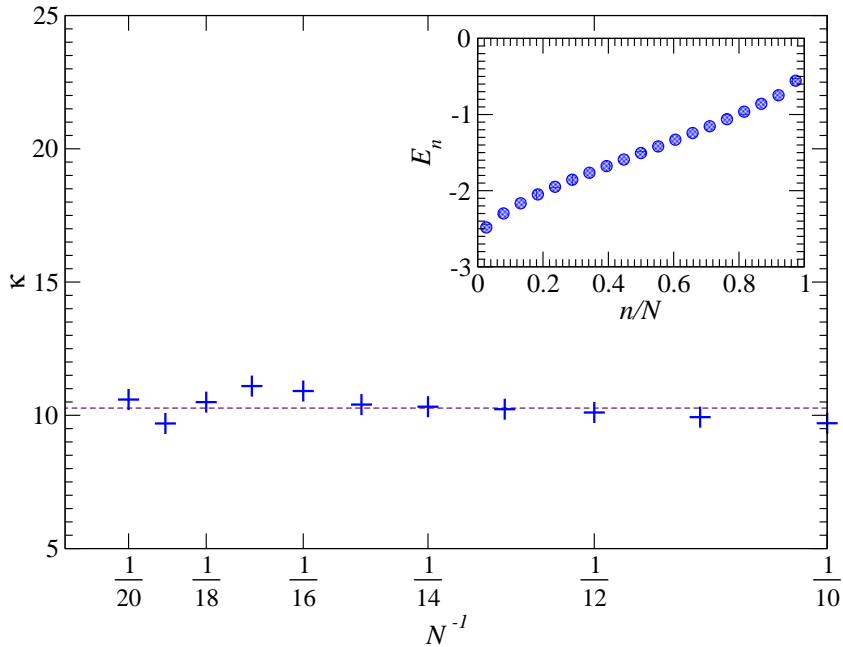}
\caption{
  Size dependence of  the energy current for the chaotic chain  with $T_L = 5$
  and $T_R  = 50$.  The  dashed line corresponds  to a $1/N$ scaling.   In the
  inset,  energy profile  computed from  the time  average of  the expectation
  value of the energy density operator $E_n = \ave{H_n}$.
\label{fig:fourier}}
\end{center}
\end{figure}

Out  of equilibrium, we  have verified  that at  high temperatures,  the local
temperature  obtained from  time  average  of the  reduced  density matrix  of
subsets  of few  spins  centered around  the  $n$-th spin  coincide with  time
averages  of the  expectation value  of the  energy density  operator $\langle
H_n\rangle$.   However, we  have observed  that out  of equilibrium  the local
temperature of the  edge spins may not coincide with the  nominal value of the
temperature of the corresponding bath. This energy jumps are commonly observed
in quantum and  classical systems. They can be understood as  the result of an
thermal resistance of the particular contact model.

Following Ref.~\cite{MMPC}, one can define a local energy current operator for
the spin model of Eq.~\ref{eq:H} as
\begin{equation} \label{eq:J}
J_{n} = h_xQ\left(\sigma_{n-1}^z-\sigma_{n+1}^z\right)\sigma^y_{n},
\quad
1\le n\le  N-2 \ ,
\end{equation}
that  is  consistent  with  the  conservation  of  energy  at  the  bulk.   In
Fig.~\ref{fig:fourier} we show the  heat conductivity $\kappa=J/\nabla T$ as a
function of  the size $N$ of  the chaotic chain  for sizes up to  $N=20$.  The
mean current  $J$ was calculated  as an average  of $\ave{J_n}$ over  time and
over the $N-8$  central spins. Three spins near each  bath have been discarded
in order  to be in the  bulk regime. For  the particular choice of  the energy
density operator (\ref{eq:H_local}), its averaged expectation value is related
to  the  local  temperature   as  $\ave{H_n}\propto  -1/T$  \cite{MMPC}.   The
temperature difference  was thus  obtained as $\Delta  T =  -1/\ave{H_{N-5}} +
1/\ave{H_3}$.   For large  $N$  the  heat conductivity  of  the chaotic  chain
converges to a  constant value, thus confirming the  validity of the Fourier's
law.   In  \cite{MMPC} it  was  also  shown that  for  the  integrable and  an
intermediate chains the Fourier's law  does not hold as $\kappa$ diverges. The
results of  Ref.~\cite{MMPC} supports the relation between  a normal transport
and the onset of quantum chaos.

\section{Quantum Master Equation for Nonequilibrium States}
\label{sec:qme}

In this  Section we  present an alternative  way of modeling  a nonequilibrium
transport scenario.   We consider  a spin-$\frac12$ chain  of length  $N$ with
homogeneous and  isotropic nearest-neighbor  Heisenberg interaction $V$  and a
contribution $H^{\text{ext}}$ due to  the interaction with an external uniform
magnetic field in $z$-direction.  The Hamiltonian of the spin chain reads
\begin{align}\label{eq:heisenbH}
        \mathcal H&=H^{\text{ext}}+V~,\\
        H^{\text{ext}}&=\sum\limits_{n=0}^{N-1}\, H_{n}=\sum\limits_{n=0}^{N-1}\frac{\Omega}{2}\,\sigma^{z}_n~,\\
        V &=\sum\limits_{n=0}^{N-2}\,V_{n,n+1}=\lambda\sum\limits_{n=0}^{N-2}\,\sigma_{n}\cdot\sigma_{n+1}~.
\end{align}
The  energy contribution due  to the  external field  of strength  $\Omega$ is
assumed to be an approximately  conserved quantity.  This is well justified if
$\Omega$ is large compared to  the coupling strength $\lambda$.  By making use
of a discretized version of the continuity equation, the following form of the
energy current operator is found
\begin{equation}\label{eq:current}
J_{n,n+1}=i\,\Kom{V_{n,n+1}}{H_{n}}~.
\end{equation}

The central ideas of the following considerations stem from the theory of open
quantum system,  commonly used to study  the properties of  systems in thermal
equilibrium, or  to account for  environmental effects in  otherwise perfectly
coherent quantum dynamics.   Our aim is to describe  the nonequilibrium steady
state, that  arises when  the chain of  spins is  in contact with  two thermal
baths at different temperatures, leading  to a steady energy current that will
flow from the hotter towards the colder heat bath through the chain of spins.

Our method  is based on  a Markovian quantum  master equation for  the reduced
density  operator $\rho_{\text S}$  of the  spin chain,  which is  designed to
model  energy transport under  a thermal  gradient. In  addition to  the usual
assumption of weak system-bath interactions, and to neglect all memory effects
(Born/Markov  approximation),  we require  the  internal interactions  between
spins to  be weak. By further assuming  not too low temperatures  of the baths
the following QME is well justified (see~\cite{Wichterich2007}, and references
therein):
\begin{align}\label{eq:wc_qme_twobath}
\dod{}{t}\rho_{\text S}(t) &=-i\Kom{\mathcal H}{\rho_{\text S}(t)} + \mathcal D_{\text L}(\rho_{\text S}(t))+\mathcal D_{\text R}(\rho_{\text S}(t))\\
\mathcal D_{\text{L}}(\rho_{\text S}(t))&= \sum\limits_{k,l=1}^{2}(\gamma_{\text{L}})_{kl}\,\Big( F_k \rho_{\text S} F^{\dagger}_l 
        - \frac{1}{2}
        [F^{\dagger}_l F_k, \rho_{\text S}]_{+}\Big)\\
F_1 &\equiv \sigma^{+}_{0},\qquad
F_2\equiv \sigma^{-}_{0}~.
\end{align}
The  dissipator of  the right  heat  bath $\mathcal  D_{\text{R}}$ is  defined
correspondingly. The coefficient matrices $\gamma_{\text{L/R}}$ read
\begin{align} \label{eq:gammas}
\gamma_{\text{L/R}} \equiv \pi\,\lambda_{\text B}\,I(\Omega)\left(
        \begin{array}{cc}
                \mathrm N(\Omega) & \sqrt{\mathrm N(\Omega)^2+\mathrm N(\Omega)}\\
                \sqrt{\mathrm N(\Omega)^2+\mathrm N(\Omega)}& \mathrm N(\Omega) + 1
        \end{array}
        \right),\quad \mathrm N(\Omega)\equiv\frac{1}{e^{\Omega\beta_{\text{L/R}}}-1}~,
\end{align}
where $\beta_{\text{L/R}}$ refers to  the reciprocal temperature of left/right
bath  respectively,  $\lambda_{\text  B}$  controls the  system-bath  coupling
strength and  $I(\Omega)$ denotes  the spectral density  of the bath,  that we
choose Ohmic.  A remarkable  property of Equation (\ref{eq:wc_qme_twobath}) is
that it can be brought into Lindblad form \cite{lindblad,gorini}
\begin{align}\label{eq:lindblad_qme}
\dod{}{t}\rho_{\text S}(t) &=-i\Kom{\mathcal H}{\rho_{\text S}(t)}+\sum\limits_{k}\alpha_k\,\Big( L_k \rho_{\text S} L^{\dagger}_k
        - \frac{1}{2} [L^{\dagger}_k L_k, \rho_{\text S}]_{+}\Big)
\end{align}
by diagonalizing  the coefficient matrices  $\gamma_{\text{L/R}}$~.  Therefore
Equation~(\ref{eq:wc_qme_twobath}) can  be treated  with the Monte  Carlo wave
function technique that will be briefly described in the following Section.

\subsection{Monte Carlo wave function method}\label{sec:unrav}

As the solution  of QME's is rarely available analytically  one is most likely
confronted  with  the  numerical  time  propagation  of  the  reduced  density
operator. Standard  methods for the numerical solution  of linear differential
equations like Runge  Kutta solvers can readily be  applied for small systems,
but fail if  the dimension of the  Hilbert space $d$ becomes large  in view of
computer memory limitations.

Since the  early 1990's the Monte  Carlo Wave Function  (MCWF) technique (also
known  as  the  quantum  jump  approach), has  become  increasingly  popular.  
Introduced in  the context of quantum  optics \cite{molmer1993,plenio1998} the
MCWF technique has also been used in quantum state diffusion unraveling of the
QME  \cite{schack} and in  the study  of the  stability of  quantum algorithms
\cite{barenco,carlo}.   The  basic  idea  is  to  depart  from  a  statistical
treatment by means of density operators  and turn to a description in terms of
stochastic wave  functions. One  might be tempted  to think of  single systems
being  continuously  monitored, but  this  analogy  is  sometimes criticized.  
Nevertheless, the  MCWF technique has proved  to be a powerful  method for the
numerical solution of Lindblad-type QME's.

The  Lindblad QME  (\ref{eq:lindblad_qme}) can  equivalently be  formulated in
terms of a stochastic Schr\" odinger equation (SSE)
\begin{align}\label{eq:sse}
 \Ket{\dd\psi} = \underbrace{\left(H_{\text{eff}}+\frac{p}{2} \right) \Ket{\psi} \dd t}_{\text{deterministic evolution}} +\underbrace{\sum\limits_{k} \left(\frac{J_k}{\sqrt{p_k}} -1\right)\Ket{\psi}\, \dd n_k}_{\text{jump-like evolution}}
\end{align}
describing     a     piecewise     deterministic    process     in     Hilbert
space~\cite{Breuer2002}, a solution of which  is is called a realization.  The
first term in Eq.~(\ref{eq:sse}) corresponds to a deterministic time-evolution
due to an effective non-Hermitian Hamiltonian given by
\begin{align}
 H_{\text{eff}}\equiv -i\mathcal H - \frac{\alpha_k}{2}\sum\limits_{k=1}L^{\dagger}_k \, L_k
\end{align}
whereas the second term  in Eq.~(\ref{eq:sse}) refers to jump-like, stochastic
evolution induced by the jump operators
\begin{align}\label{eq:jumpops}
 J_k\equiv\sqrt{\alpha_k}\,L_k~.
\end{align}
The  Poisson  increments $\dd  n_k\in\lbrace  0,1\rbrace$  obey the  following
statistical properties
\begin{align}
 \expec{\dd n_k} &= p_k\, \dd t~,\\
 \dd n_k\,\dd n_l&=\delta_{kl}\,\dd n_k~.
\end{align}
$\expec{\cdot}$ stands for the expectation value, whereas $p_k$ denotes a jump
rate given by
\begin{equation}
p_k = \Vert\,J_k\Ket{\psi}\Vert^2 \ ,
\end{equation}
and  is therefore time-dependent.   $p=\sum_k\,p_k$ refers  to the  total jump
rate.  The SSE~(\ref{eq:sse})  is an  equivalent formulation  of  the Lindblad
QME~(\ref{eq:lindblad_qme}) insofar as
\begin{equation}
 \expec{\Ket{\psi(t)}\Bra{\psi(t)}}=\rho_{\text S}(t) \ ,
\end{equation}
given that  $\expec{\Ket{\psi(t_0)}\Bra{\psi(t_0)}}=\rho_{\text S}(t_0)$. Thus
the  expectation value  of an  observable  $A$ at  time $t$  can be  estimated
through
\begin{equation}
 \langle A\rangle\,(t)=\Tr\lbrace A\,\rho_{\text S}(t)\rbrace\simeq\frac{1}{m}\sum\limits_{k=1}^{m}\Bra{\psi_k(t)}A\Ket{\psi_k(t)}
\end{equation}
in  a  finite ensemble  of  $m$  realizations  of~(\ref{eq:sse}) to  arbitrary
precision.  This is  of  huge practical  importance,  as one  deals with  wave
functions with  $\mathcal O(d)$ elements  instead of density  operators having
$\mathcal O(d^2)$  elements, where $d$ is  the dimension of  the Hilbert space
under  consideration. Furthermore,  if  one is  interested  in the  stationary
state,  ensemble averages  can be  replaced by  time averages  and  one single
realization    suffices   to    determine   stationary    expectation   values
\cite{molmer1996,cresser}
\begin{equation}\label{eq:timeav}
 \langle A\rangle^{\text{stat}}=\Tr\lbrace A\,\rho_{\text S}^{\text{stat}}\rbrace\simeq\frac{1}{(T+1)}\sum\limits_{k=0}^T \Bra{\psi (t_k)} A\Ket{\psi (t_k)},\quad t_k \equiv t_0 + k\, \Delta t~.
\end{equation}

The MCWF approximation to the exact  solution depends on the number of quantum
trajectories  that are considered,  as well  as the  integration time  of each
trajectory.  We  have found  that one single  trajectory, integrated  during a
sufficiently long  time, results in a very  good approximation.  Nevertheless,
following~\cite{molmer1996}, we consider $i=1\ldots m$ different realizations,
so  that, an estimate  of $\Tr\lbrace  A\,\rho_{\text S}^{\text{stat}}\rbrace$
can    be    obtained    as     the    sample    mean    of    the    $\langle
A\rangle^{\text{stat}}_i$'s.     Moreover,    considering   several    quantum
trajectories  enables  us   to  get  a  measure  for   the  statistical  error
$\delta\langle    A\rangle^{\text{stat}}/\sqrt{m}$,    where    $\delta\langle
A\rangle^{\text{stat}}$   is   the   standard   deviation  of   the   $\langle
A\rangle^{\text{stat}}_i$'s~.  The results  presented below were obtained from
$m=4$ different realizations, integrated for long times of the order of $10^5$
to $10^7$.

We  now   describe  the  procedure  to   obtain  a  realization   of  the  SSE
(\ref{eq:sse}).

\noindent Starting from an initial state $\Ket{\psi(t_0)}$ we proceed as follows:
\begin{enumerate}
\item Draw a uniformly distributed random number $r\in [0,1]$.
\item        Perform       the       deterministic        time       evolution
  $\mid\tilde{\psi}(t)\,\rangle=e^{(t-t_0)      H_{\text{eff}}}\Ket{\psi(t_0)}$
  until $t=t_j$, which is determined by $\Vert\tilde{\psi}(t_j)\Vert^2=r$, for
  some $r<1$.
\item Normalize the wave function $\Ket{\psi}\rightarrow\,\mid\tilde{\psi}\,\rangle\,/\,\Vert\tilde{\psi}\Vert$.
\item Choose randomly a particular jump $k$ with respective weight $p_k$.
\item Carry out the map $\Ket{\psi}\rightarrow\,J_k\Ket{\psi}\,/\,\sqrt{p_k}$.
\item Set $t_0\rightarrow t$ and return to step 1.~
\end{enumerate}
This procedure  terminates when $t=t_{\text{fin}}$,  where $t_{\text{fin}}$ is
the desired final time.

The jumps occur at random instants of time $t_j$, which are determined through
$\Vert\tilde{\psi}(t_j)\Vert^2=r$,  by  virtue  of  the  second  step  of  the
simulation procedure. Assuming  a uniform time discretization $\Delta  t$ in a
simulation,  jumps happen only  at multiples  of $\Delta  t$, which  causes an
error of order  $\mathcal O(\Delta t)$. Therefore $\Delta t$  has to be chosen
with  care. For a  more detailed  account on  the MCWF  method, the  reader is
referred to~\cite{Breuer2002}~.
% By simulating a finite number of stochastic wave functions, expectation values can thus be estimated to arbitrary precision.
% The advantage of using wave functions instead of density operators is the number of independent elements, which is of $\mathcal O(N^2)$ in the latter, but only $\mathcal O(N)$ in the former case.

% One recognizes that the stochastic description, which was introduced above
% has several similarities to the type of bath modeling that was presented
% earlier in this work.

\subsection{Fourier's Law in the Heisenberg chain}
\label{sec:results}

\begin{figure}[!t]
\begin{center}
\includegraphics[scale=2.0]{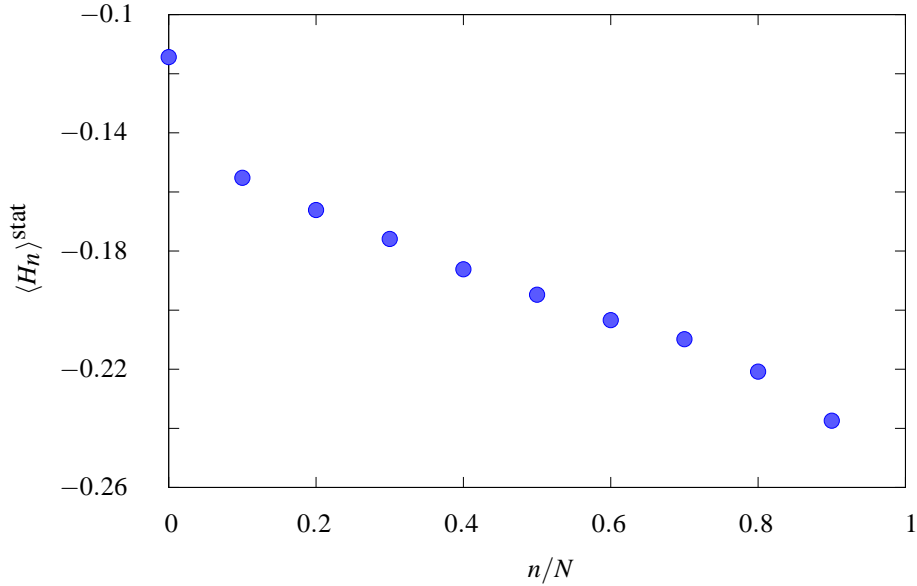}
\caption{
  Energy profile in  a Heisenberg  chain of  $N=10$  spins-$\frac12$~. A
  finite  gradient  is   associated  with   diffusive  transport
  behaviour.  The data  points refer  to  time averages  of single  stochastic
  realizations  up   to  a  final  time  $t_{\text{fin}}=10^5$   in  units  of
  $\Omega^{-1}$~.  System  parameters:  ($\beta_{\text  L}=0.41,\,\beta_{\text
    R}=1.39,\,\lambda=\lambda_{\text B}=0.01,\,\Omega=1$)~.
\label{fig:tprofile}}
\end{center}
\end{figure}

We consider  the Heisenberg chain of spin-$\frac12$  with Zeeman contribution,
described  by  the  Hamiltonian  (\ref{eq:heisenbH}),  as  an  example  of  an
integrable chain.  As we  have mentioned, integrability is commonly associated
with diverging transport coefficients  and hence ballistic transport behavior. 
However, in Ref.~\cite{MHGM}, normal transport was observed for an integrable,
albeit  small, spin  chain.   In Fig.~\ref{fig:tprofile}  we  show the  energy
profile along  the Heisenberg  chain in the  stationary state. A  clean finite
temperature gradient is observed.

\begin{figure}[!t]
\begin{center}
\includegraphics[scale=2.0]{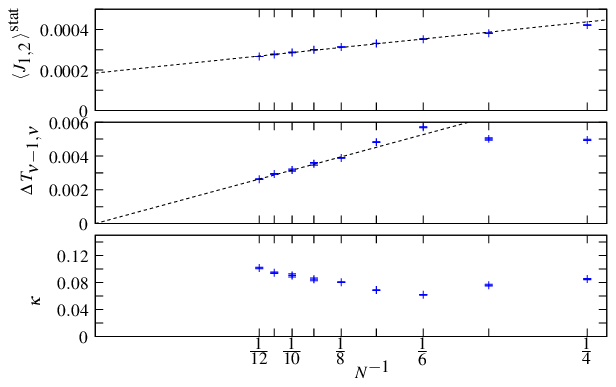}
\caption{
  Scaling behavior of heat conductivity.  The different panels show the energy
  current  (top),  the temperature  gradient  (middle)  and heat  conductivity
  $\kappa$ (bottom)  in the stationary state  as a function  of the reciprocal
  chain length $N^{-1}$.   The results where obtained by  the MCWF method.  In
  the upper  and middle panels  the dashed line  corresponds to a  linear best
  fit. In  the middle panel only data  points with $N\geq 8$  were considered. 
  The error bars  in $\kappa$ where obtained from  error propagation in linear
  approximation.                       System                      parameters:
  $\beta_L=0.25,\,\beta_R=0.5,\,\lambda=\lambda_{\text B}=0.01,\,\Omega = 1~.$
  Simulation   parameters:   $\Delta   t  =   1,\,t_{\text{fin}}=10^7   (N\geq
  9),\,t_{\text{fin}}=10^8 (N\leq 8)~.$
\label{fig:1}}
\end{center}
\end{figure}

We explicitly compute transport coefficients  here for Heisenberg chains of up
to 12  spins, by making use of  the MCWF technique, taking  time averages over
single trajectories. We focus on the scaling behavior of the heat conductivity
$\kappa$ with the  size of the chain  $N$. We define $\kappa$ as  the ratio of
two measures, namely the stationary  energy current within the chain, in terms
of  the current operator  given in  Eq.~(\ref{eq:current}) and  the stationary
difference in the local energy of the innermost pair of spins
\begin{align}\label{eq:kappa}
 \kappa&\equiv\frac{\langle J_{1,2}\rangle^{\text{stat}}}{\Delta\, T_{\nu-1,\nu}}~,\\
\Delta\, T_{\nu-1,\nu}&\equiv\langle\,H_{\nu-1}-H_{\nu}\,\rangle^{\text{stat}},\quad\nu\equiv\frac{N}{2}-\text{mod}(N,2)~.
\end{align}
In the upper panel of Fig.~\ref{fig:1} it is shown that the stationary current
depends linearly on the reciprocal chain length $N^{-1}$.  The same result has
been observed earlier  in Ref.~\cite{MHGM} on the basis of  a similar QME, but
for chain  lengths of $N\leq  6$.  Extrapolation of  a linear best fit  of the
data points in the upper  panel of Fig.~\ref{fig:1} strongly suggests that, in
the limit  of infinite $N$ the  stationary energy current  remains finite.  In
contrast  to the  diffusive  behaviour  observed for  small  systems, this  is
expected to wane at the thermodynamic limit $N\rightarrow\infty$ .

In a  real material the  heat conductivity $\kappa$  is a bulk  property. When
transport  is normal,  $\kappa$ converges  towards  a constant  value with  an
increasing  size of  the  system, once  finite  size effects  are negligible.  
However, our results  show no sign of a convergence for  $\kappa$ in the range
of   our   computational   abilities.    Linear  extrapolation   of   $\langle
J_{1,2}\rangle^{\text{stat}}$ and  $\Delta\, T_{\nu-1,\nu}$ rather  predicts a
divergence of the  so defined heat conductivity with $N$.   This agrees with a
ballistic transport in the integrable systems.

\section{Discussion}
\label{sec:disc}

In  the  previous sections  we  have described  two  formalisms  to study  the
dynamics of  quantum spin  chain models in  thermal nonequilibrium  states. In
this section we discuss their limitations and applicability .

Both methods successfully generate a  nonequilibrium state in the bulk of spin
chain  models. In the  stochastic baths  method the  information of  the model
enters  in  the  precise  definition  of the  local  measurement  periodically
performed on the spins in contact with the baths. For the QME, the particulars
of the system are defined in the decay rates of Eq.~\ref{eq:gammas}.

However,  the two  methods have  a very  different character.   The derivation
presented in Section~\ref{sec:qme} obtains a QME of Lindblad type, appropriate
to describe  a quantum state for which  the temperature field is  not uniform. 
The fact  that Eq.   \ref{eq:wc_qme_twobath} can be  written in  Lindblad form
makes possible  to compute averages  of thermodynamical observables  using the
Monte  Carlo  wave  function  formalism,  in  which  each  quantum  trajectory
corresponds to a stochastic unraveling of the QME (\ref{eq:wc_qme_twobath}).

The model  of Section~\ref{sec:sto}  positively neglects any  particular model
for the heat baths and defines  a procedure by which the system stochastically
dissipates at its boundaries.  As so, this formalism does not corresponds to a
stochastic unraveling  of a master equation  for the density  matrix operator. 
The  evolution of  the system  between two  consecutive interactions  with the
baths is  purely unitary and not  continuously dissipative as in  the QME. The
stochastic method is an approximation of the QME.

The  use of  the MCWF  technique to  obtain averaged  expectation  values make
possible to numerically study larger  systems than with other methods. Without
this, the solution of  Eq.~\ref{eq:wc_qme_twobath} involves the integration in
full Liouville space of  dimension $2^{2N}$, limiting numerical investigations
to small system  sizes (typically of $N\le  6$). The price to pay  is that the
Lindblad  QME  (\ref{eq:wc_qme_twobath}), is  valid  only  for weakly  coupled
spins.   This  limitation is  particularly  relevant  for  chaotic systems  as
typically,  chaotic  behaviour  is  exhibited  above  a  critical  interaction
strength.  Despite its limitations,  Eq.~\ref{eq:wc_qme_twobath} is, up to our
knowledge, the  only rigurous QME  of Lindblad form  that is adapted  to study
systems in a nonequilibrium state.

The stochastic baths method does  not present this limitation. The strength of
the spin  interaction determines the internal  relaxation time of  the system. 
Since the  frequency $\tau^{-1}$  at which the  system dissipates in  the heat
baths is a free parameter, one  can always find an appropriate value of $\tau$
for which, the stationary  nonequilibrium state is established.  Moreover, the
stochastic baths can be generalized  to consider more general situations, like
{\em e.g.}   the coupling with thermo-magnetic baths,  for which thermodynamic
cross effects can be studied.  Nevertheless, the lack of a model for the baths
limits a precise interpretation of the physical dissipation.

\section{Conclusions}
\label{sec:final}

We have presented two complementary methods to study heat transport in quantum
spin chains. The first is based  on a stochastic procedure by which, the state
of the  subsystems that are coupled to  ideal heat baths is  consistent with a
global nonequilibrium state. The second is based on a QME in Lindblad form.

The stochastic baths method does only require the integration of the pure wave
function  of the system.   The Lindblad  QME (\ref{eq:wc_qme_twobath})  can be
integrated by means  of Monte Carlo wave function  techniques.  Therefore, the
use of any  of these methods allows  us to study longer spin  chains than with
other methods.  This  is particularly relevant to the  study of nonequilibrium
states as the quantities that  determine the transport properties are formally
defined in the limit of infinite volume.

We have  shown the application of these  two methods to study  the validity of
the quantum Fourier's  law in a chaotic and an integrable  quantum spin chain. 
As generally observed,  we have obtained that the Fourier's  law holds for the
chaotic chain, while for the  integrable chain, the heat conductivity diverges
with the size of the chain.

It would be  interesting to compare the nonequilibrium  state generated by the
two methods  presented here.  An investigation in  this direction  will appear
elsewhere \cite{future}.

\begin{acknowledgement}
  C.M.-M. acknowledge a Lagrange  fellowship from the Institute for Scientific
  Interchange Foundation.
\end{acknowledgement}

\end{document}